# Few-Shot Learning Approach on Tuberculosis Classification Based on Chest X-Ray Images

1st A.A.G. Yogi Pramana
*Department of Computer Science and Electronics*
*Universitas Gadjah Mada*
Yogyakarta, Indonesia
aagdeyogipramana@mail.ugm.ac.id

2nd Faiz Ihza Permana
*Department of Electrical Engineering and Information Technology*
*Universitas Gadjah Mada*
Yogyakarta, Indonesia
faiz.ihza.permana@mail.ugm.ac.id

3rd Muhammad Fazil Maulana
*Department of Computer Science and Electronics*
*Universitas Gadjah Mada*
Yogyakarta, Indonesia
muhammadfazilmaulana@mail.ugm.ac.id

4th Dzikri Rahadian Fudholi
*Department of Computer Science and Electronics*
*Universitas Gadjah Mada*
Yogyakarta, Indonesia
dzikri.r.f@ugm.ac.id

*Abstract*—Tuberculosis (TB) is caused by the bacterium *Mycobacterium tuberculosis*, primarily affecting the lungs. Early detection is crucial for improving treatment effectiveness and reducing transmission risk. Artificial intelligence (AI), particularly through image classification of chest X-rays, can assist in TB detection. However, class imbalance in TB chest X-ray datasets presents a challenge for accurate classification. In this paper, we propose a few-shot learning (FSL) approach using the Prototypical Network algorithm to address this issue. We compare the performance of ResNet-18, ResNet-50, and VGG16 in feature extraction from the TBX11K Chest X-ray dataset. Experimental results demonstrate classification accuracies of 98.93% for ResNet-18, 98.60% for ResNet-50, and 33.33% for VGG16. These findings indicate that the proposed method outperforms others in mitigating data imbalance, which is particularly beneficial for disease classification applications.

*Index Terms*—TBX11K, Prototypical Network, Data Imbalance, Transfer Learning

## I. INTRODUCTION

### A. Research Background

Tuberculosis (TB) is an infectious disease caused by the bacterium *Mycobacterium tuberculosis* and is the leading cause of death from a single infectious disease [1]. The bacteria commonly invade the lungs leading to reduced efficiency of lung function. It can also cause damage to other parts of the body such as the brain or spine [2]. The potential spread of TB can be very high through the medium of air when people with TB cough and sneeze.

Early diagnosis of TB can help increase the chances of a person with TB to recover. Early diagnosis of TB means that treatment can be initiated earlier, leading to less chance of transmission and better outcomes for patients [3]. One of the ways for early diagnosis of TB is by using image analysis with X-ray modality, especially on the lungs. X-ray images can reveal several findings that may indicate the presence of TB, with symptoms such as nodules, cavities, or infiltrates in the patient's lung tissue.

TB diagnosis can be categorised into active and latent types. Active TB infection is risky and this type of TB is distinguished by consolidation of cavitary lesions in the lungs, has a strong likelihood of spreading the disease [4], while latent TB infection is the presence of immunoreactivity to tuberculosis antigens in the absence of clinical and radiological symptoms of TB. However, latent TB infection can reactivate and cause infectious disease [5]. In addition, there are also cases of active and latent TB. In this case, both active and latent TB can be seen in an X-ray image. This is a rare case, where the rarity of the case causes data imbalance. Data imbalance can be resolved by data augmentation such as zooming, flipping, and rotating the chest X-ray image for TB detection [6].

However, data augmentation can produce variations that do not reflect the original conditions, which can make the model evaluation experience overfitting. The prototypical network algorithm based on the few-shot learning (FSL) approach can overcome this problem. The working principle of this algorithm is that the classifier should generalise to new classes not seen in the training set, with only a small number of examples of each new class through metric space learning where classification can be performed by calculating the distance to the prototype representation of each class with a simpler inductive bias [7].

Therefore, a FSL approach through prototypical network algorithm is used to handle the data imbalance problem of active-latent tuberculosis disease. The method was tested using the TBX11K dataset (https://www.kaggle.com/datasets/usmanshams/tbx-11) which contains patient X-ray image data. This dataset contains data of 3800 TB negative patients and 800 TB positive patients, with the percentage of TB positive patients reaching 17% of the total data.

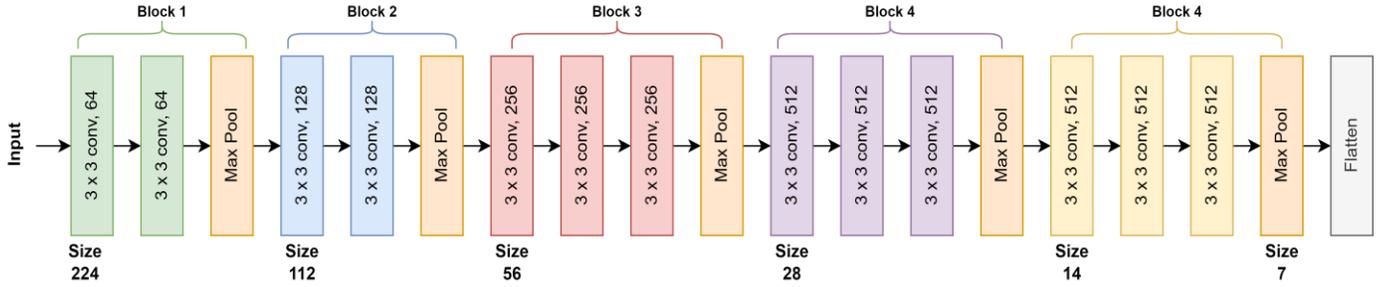

Fig. 1. Modified VGG-16 Architecture Without Fully Connected Layer

*B. Problem Definition*

This research focuses on addressing the issue of class imbalance within the TBX11K chest X-ray image dataset, a common challenge encountered in medical imaging datasets. In particular, the dataset exhibits a significant disproportion between the negative TB class, which is substantially larger, and the positive TB class. This imbalance poses a challenge for classification models, as they tend to prioritize learning from the majority class, ultimately leading to reduced accuracy in predicting instances of the minority class.

*C. Research Constraints*

In this study, several problem constraints are provided as follows:
1) The TBX11K dataset, obtained from Kaggle, serves as the primary dataset for analysis.
2) mage feature extraction is performed using three distinct models: VGG16, ResNet-18, and ResNet-50.
3) The classification task is implemented using the Prototypical Learning algorithm.
4) The objective of the study is to examine the impact of varying shot numbers and feature extraction methods on classification performance in the context of imbalanced image data.

*D. Research Contributions*

The literature review indicates that previous research has addressed the challenge of data imbalance through the application of various data sampling techniques and classification methods, including Random Forest and XGBoost, as highlighted in the surveyed studies. However, there is a notable gap in the literature concerning the use of a few-shot learning approach, specifically Prototypical Networks, to address class imbalance in chest X-ray (CXR) image datasets.

## II. LITERATURE REVIEW

Data imbalance is common in medical machine learning, impacting model accuracy in disease diagnosis. One approach to address this is data augmentation, involving techniques like zooming, flipping, and rotating, as used by Oltu et al. [6] for TB detection in chest X-rays, achieving 96.6% accuracy and a 0.99 AUC. Similar augmentation methods were applied by Kotei [8] and Bista [9], preventing overfitting. However, data augmentation risks generating unrealistic variations, leading to overly optimistic evaluations.

Another method, SMOTE, addresses data imbalance by oversampling minority classes. Fadhlullah [10] used SMOTE on the TBX11K dataset, improving accuracy from 94.11% to 94.33% with XGBoost. Similarly, Fonseca [11] achieved 99.74% accuracy using SMOTE. However, SMOTE can introduce issues such as data overlap, noise, and small disjuncts [12] [13].

To overcome the weaknesses of those methods, the use of prototypical network algorithm, which is a few-shot learning approach, is considered. This algorithm proposed in [7] is used to solve the problem of few-shot classification. This algorithm is able to cover the weaknesses in other techniques so it can solve the problems of few-shot learning, especially overfitting. Thus, this study aims to explore the impact of prototypical network algorithm which is a few-shot learning approach on imbalanced chest X-ray image dataset. This research is expected to provide a comparison of model prediction results with and without the use of prototypical network algorithms.

## III. METHODOLOGY

*A. Research Description*

This research aims to support neurosurgeons in more accurately diagnosing tuberculosis types by addressing the data imbalance issue in TB X-ray images without resorting to oversampling methods. By integrating meta-learning and machine learning, the proposed classification model enhances the precision of distinguishing between TB classes, crucial for improving diagnostic accuracy and treatment decisions.

The success of classification models depends heavily on data volume and distribution across classes [14]. In imbalanced datasets, a high overall accuracy often masks poor performance in minority classes, leading to misclassification of critical cases. To tackle this, few-shot learning offers a robust solution. Designed for scenarios with limited data, it leverages knowledge from related tasks and advanced techniques like metric and transfer learning to improve minority class representation. This approach not only enhances classification accuracy for underrepresented classes but also improves overall model performance, making it particularly effective for real-world imbalanced datasets.

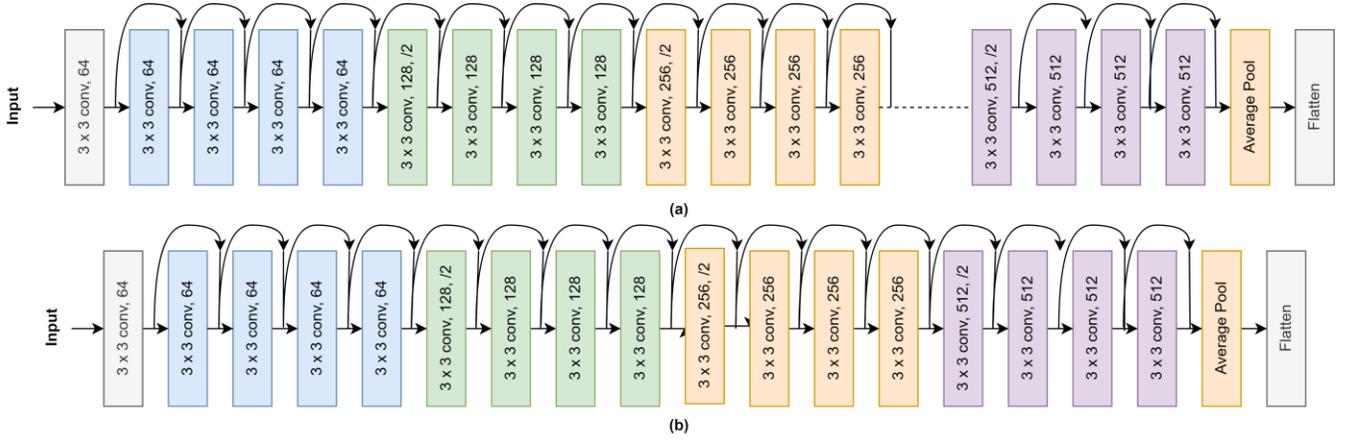

Fig. 2. Modified ResNet-50 (a) and ResNet-18 (b) Architecture Without Fully Connected Layer

### B. Data Acquisition

The second stage of this research is the data acquisition stage. The data used in this study is the TBX11K dataset, which is a public dataset available on Kaggle. TBX11K consists of chest X-ray (CXR) images of TB patients from various sources. In this study, the TBX11K dataset elaborated by [15] was used, where each image is annotated with bounding boxes for classifying the detected type of TB. The image resolution in this dataset is 512×512. The dataset comprises a total of 11,200 CXR images with a strong bias to the number of Healthy and Sick non-TB X-ray images. Therefore, before modelling, it is necessary to apply a data balancing technique on the training data to ensure that the number of samples in each class is treated as balanced.

### C. Algorithm Overview

The next stage is modelling. The model in this study are divided into two. First, model without training, as shown in Figure 3. In this model, the test image is directly tested to the model without training. In the other hand, in model with training, Figure 4, the dataset train is being trained in the first place, then the test train is tested after the training process is done. In this third stage, the processes involved include resizing and feature extraction using different backbones for the prototypical algorithm. To address the imbalanced dataset, which includes only three classes (TB, Sick, Healthy), a few-shot learning approach was utilized with the Prototypical Network algorithm. This algorithm required a backbone, so ResNet18, ResNet50, and VGG16 were employed as varying backbones. Each backbone was tested across four different variations: 3-Way 20-Shot, 3-Way 10-Shot, 3-Way 5-Shot, and 3-Way 1-Shot. All variations were evaluated using a query set of 10. samples. This methodology aimed to improve the model's ability to generalize across the different classes despite the limited data available for some categories.

### D. Preprocessing Model

At this phase, the image data undergoes preprocessing in preparation for feature extraction utilizing the VGG16,

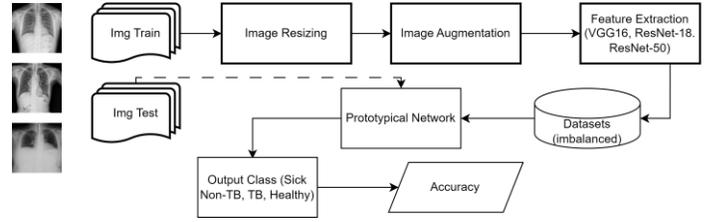

Fig. 3. Model Without Training

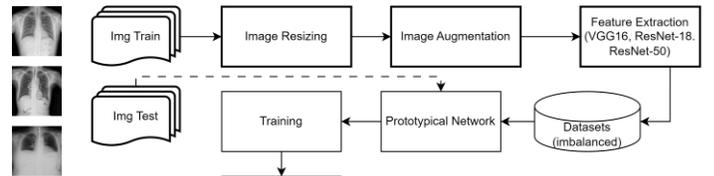

Fig. 4. Model Without Training

ResNet-18, and ResNet-50 architectures. The preprocessing pipeline involves resizing the images to 224x224 pixels and transforming the color space to RGB format. Furthermore, data augmentation methods, including cropping, flipping, and rotation, are implemented to increase the variability of the dataset for improved model robustness.

### E. Feature Extraction

Feature extraction is performed using three pre-trained CNN models: VGG16, ResNet-18, and ResNet-50. VGG16 (Figure 1) consists of 16 layers, ResNet-18 (Figure 2b) has 18 layers, and ResNet-50 (Figure 2a) contains 50 layers. For this study, fully connected layers and dropout are removed from all models, as they are used solely for feature extraction, not classification. Specifically, the last 3 layers of VGG16 and the last 2 layers of ResNet-18 and ResNet-50 are omitted.

VGG16, ResNet-18, and ResNet-50 all utilize pre-trained convolutional layers from the ImageNet dataset to facilitate transfer learning. In all three models, the convolutional layers are frozen, preserving their pre-trained weights, which significantly reduces training time and computational demands. This enables VGG16, ResNet-18, and ResNet-50 to function as feature extractors for chest X-ray (CXR) image recognition tasks.

The features extracted by VGG16, ResNet-18, and ResNet-50 are then prepared for classification using the Prototypical Network. To ensure compatibility, the extracted features are flattened into one-dimensional vectors, making them suitable for calculating Euclidean distances within the Prototypical Network.

*F. Prototypical Network*

One of the most well-liked and successful methods in the few-shot learning literature is prototypical networks [7]. Prototypical Networks were developed as an answer to the enduring problem of few-shot learning overfitting. Prototypical Networks use non-linear mappings to embed spaces, working under the premise that classifiers ought to possess a straightforward inductive bias [16]. Prototypical networks compute an $M$-dimensional representation $c_k \in R^M$ (or *prototype*) of each class via an embedding function $f_\phi : R^D \to R^M$ with learnable parameters $\phi$ [7]. For Each prototype is the mean vector of the embedded support points that belong to its class:

$$c_k = \frac{1}{|S_k|} \sum_{(x_i, y_i) \in S_k} f_\phi(x_i) \quad (1)$$

$S_k$ denotes the set of examples labelled with class $k$, with a supporting set of $N$ labelled examples $S = \{(x_1, y_1), ..., (x_N, y_N)\}$ where each $x_i \in R^D$ is a $D$-dimensional feature vector of an example and $y_i \in \{1, ..., K\}$ is the corresponding label. As illustrated in Figure 5, this approach is simpler and more efficient than many other meta-learning algorithms, making it an attractive solution for addressing few-shot learning tasks. [17].

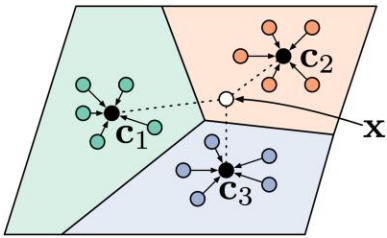

Fig. 5. Prototypical Networks applied in few-shot scenarios. Few-shot prototypes, denoted as $c_k$ are determined as the average of embedded support examples for each class [7]

Few-shot learning is essential in medical image classification, due to limited labeled data, adapting to new imaging modalities, and reducing the anotation burden [16]. To generate a prototype vector for each class, the vectors of the N examples from that class are concatenated. The test data point is then classified by calculating its distance from the prototype vectors of each class. [17]. To apply the Prototypical Network to image data, the images are first processed using three different pre-trained CNN models—VGG16, ResNet-18, and ResNet-50. These models extract image features and convert them into one-dimensional vectors, which represent the feature embeddings of the images.

*G. Classification Model*

The classification process employs a Prototypical Network for few-shot learning, leveraging feature vectors extracted from VGG16, ResNet-18, and ResNet-50. The network operates with a support set for training and a query set for testing, aiming to classify query images into three categories: TB, Sick, and Health.

Class prototypes are generated by averaging the feature vectors of support set examples for each class. For instance, the prototype for the "TB" class is the mean of the feature vectors from its support examples, with the same procedure applied for Sick and Health classes. Query image classification is performed by comparing the query's feature vector, extracted by the CNNs, to the class prototypes using Euclidean distance. The query image is assigned to the class with the closest prototype. Model performance is then evaluated by comparing the predicted labels with the true labels, with accuracy used as the key metric for assessing classification effectiveness across the three classes.

*H. Model Testing*

*1) Validation Technique:* In the model creation process, the training data is divided with a composition of 80:20, with 80 percent allocated for training and 20 percent for model validation. This division aims to measure the model's performance during the training and classification process.

*2) Model Evaluation:* After validation, testing is conducted using a confusion matrix to evaluate the model's ability to classify data into each class. The model's performance is tested both with and without FSL resampling to compare outcomes. The dataset is split into training and testing sets, where the training data is used to develop the model, and the testing data is used for an objective performance evaluation.

Model evaluation includes calculating accuracy as a weighted average, accounting for the ratio of test samples in each class. These evaluation metrics are then used to compare the overall performance of each model.

## IV. RESULT AND DISCUSSION

*A. Preprocessing and Feature Extraction Results*

Data preprocessing was performed to prepare the dataset for feature extraction. This process involved normalizing the pixel values by dividing each value by 255, resulting in a standardized range between 0 and 1. This normalization enhances stability and improves the convergence of the training process. Additionally, all images were resized to 224x224 pixels to align with the input size requirements of the VGG16 architecture. Figure 6 illustrates an example of the resized images, showing samples from the Active and Latent TB classes.

TABLE I
RESULT OF PROTONET CLASSIFIER

| Backbone | Shot | Without Training | | With Training | |
|---|---|---|---|---|---|
| | | Accuracy | Time | Accuracy | Time |
| VGG16 | 1 | 48.30% | 00 min 12 s | 33.33% | 01 h 47 min |
| | 5 | 64.43% | 02 min 57 s | 33.33% | 05 h 30 min |
| | 10 | 69.43% | 04 min 00 s | 33.33% | 03 h 24 min |
| | 20 | 71.77% | 00 min 34 s | 33.33% | 04 h 50 min |
| ResNet-18 | 1 | 55.53% | 04 min 10 s | 96.80% | 02 h 06 min |
| | 5 | 68.90% | 02 min 09 s | 97.57% | 02 h 36 min |
| | 10 | 72.30% | 02 min 47 s | 97.80% | 03 h 22 min |
| | 20 | 74.80% | 02 min 43 s | 98.93% | 04 h 54 min |
| ResNet-50 | 1 | 51.60% | 04 min 12 s | 95.50% | 02 h 34 min |
| | 5 | 64.17% | 02 min 14 s | 96.93% | 02 h 40 min |
| | 10 | 68.47% | 04 min 31 s | 97.30% | 03 h 33 min |
| | 20 | 73.43% | 03 min 41 s | 98.60% | 05 h 04 min |

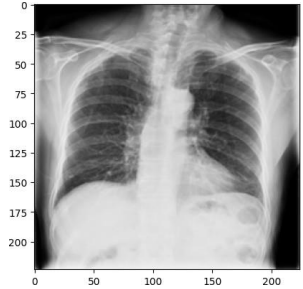

Fig. 6. Example of Resized Image Data

During the feature extraction phase, features from both the training and testing data were extracted using the VGG16, ResNet-18, and ResNet-50 models, excluding the fully connected layer (FCL). The FCL was omitted since its purpose is to classify data based on extracted features, whereas this study employs a separate classification model—the Prototypical Networks algorithm. Consequently, the output dimensions were (1, 512, 7, 7) for both VGG16 and ResNet-18, and (1, 2048, 7, 7) for ResNet-50. These outputs represent the numerical feature maps generated by processing the images through the respective CNN layers. This process is depicted in Figure 7.

```
# Output the feature map sizes
print("VGG16 feature map size:", vgg16_features.shape)
print("ResNet-18 feature map size:", resnet18_features.shape)
print("ResNet-50 feature map size:", resnet50_features.shape)

VGG16 feature map size: torch.Size([1, 512, 7, 7])
ResNet-18 feature map size: torch.Size([1, 512, 7, 7])
ResNet-50 feature map size: torch.Size([1, 2048, 7, 7])
```

Fig. 7. Output Dimension of Feature Extraction Results

### B. Model Results

The results from the three graphs clearly highlight distinct behaviors of the VGG16, ResNet-18, and ResNet-50 backbones in few-shot learning, both with and without meta-training. This discussion aims to explore the trends observed in the accuracy performance and offer possible insights into the underlying mechanisms.

In the VGG16 backbone, Figure 8, without meta-training, the accuracy increases significantly as the number of shots increases. Starting from 48.30% with 1 shot, it reaches 71.77% with 20 shots. This trend indicates that VGG16, a relatively older architecture, benefits greatly from additional data, as the number of support samples per class increases.

However, the VGG16 results with meta-training are remarkably different. The accuracy remains constant at 33.33% across all shots, suggesting that meta-training fails to adaptively improve the model's performance. This might be due to the architectural limitations of VGG16, as it is known to have fewer representational capabilities compared to more modern architectures like ResNet. This could also imply that the meta-training process may not be effectively utilizing the inductive bias introduced by the VGG16 feature space, or the optimization is stuck in a poor local minima during meta-training.

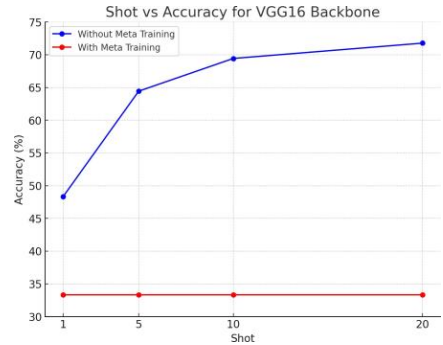

Fig. 8. The graph shows the accuracy of the VGG16 backbone in a few-shot learning task. Without meta training, accuracy improves with the number of shots, reaching around 72% at 20 shots. With meta training, the accuracy remains constant at 33.33% across all shots.

ResNet-18, a lighter model in the ResNet family, demonstrates significant accuracy gains without meta-training, improving from 55.53% with 1 shot to 74.80% with 20 shots. Its residual connections enhance generalization even with few samples, and its computational efficiency makes it ideal for few-shot learning without meta-training.

With meta-training, ResNet-18 achieves 96.80% accuracy with just 1 shot, reaching 98.93% with 20 shots. This suggests that meta-training enables ResNet-18 to quickly adapt to new tasks, with minimal accuracy gains from additional data, as the model generalizes effectively after seeing only a few examples.

ResNet-50 backbone, Figure 10, being a deeper and more complex architecture compared to ResNet-18, shows strong performance in both meta-trained and non-meta-trained scenarios. Without meta-training, the accuracy increases from 51.60% with 1 shot to 73.43% with 20 shots. Similar to ResNet-18, ResNet-50 shows a consistent improvement in performance as more examples are available, but the higher starting point suggests that ResNet-50 has better representa-

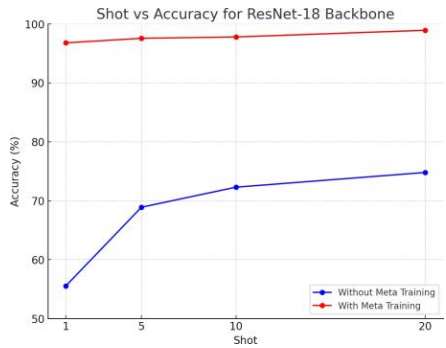

Fig. 9. The ResNet-18 backbone exhibits increasing accuracy without meta training as the number of shots increases, reaching 74.80% at 20 shots. With meta training, the accuracy starts high at 96.80% with just one shot and rises to nearly 99% at 20 shots.

tional capacity even without meta-training. With meta-training,

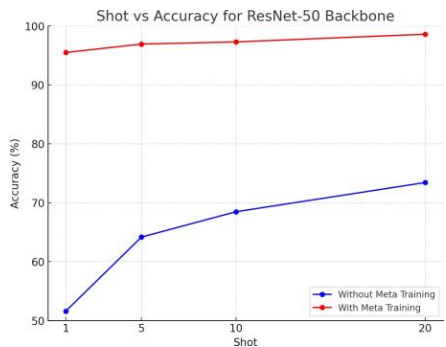

Fig. 10. The graph for ResNet-50 shows an improvement in accuracy without meta training, starting at 51.60% with one shot and reaching 73.43% at 20 shots. With meta training, accuracy starts high at 95.50% and gradually increases to 98.60% at 20 shots.

ResNet-50 demonstrates high starting accuracy (95.50% with 1 shot) and continues to increase, reaching 98.60% at 20 shots. This is similar to ResNet-18 but with a slight edge in performance, likely due to its deeper layers, which can capture more abstract features. The diminishing increase in accuracy with more shots suggests that ResNet-50, when meta-trained, has already generalized well with just a few examples, and adding more examples provides minimal additional benefit.

## V. CONCLUSION

The comparative analysis of VGG16, ResNet-18, and ResNet-50 in FSL tasks highlights the significance of both model architecture and meta-training in addressing imbalanced datasets in TB X-ray images. While VGG16 shows limitations in adapting during meta-training, ResNet-18 and ResNet-50 deliver superior performance, particularly when integrated with meta-learning. Meta-training enhances the model's ability to generalize from minimal data, leading to improved classification accuracy for underrepresented TB classes. This underscores that deep architectures with residual connections, combined with meta-learning, are the most effective strategy for rapidly adapting to new tasks with limited data, aligning with the aim of developing a reliable TB diagnosis model.